\documentclass[final,5p,times,twocolumn]{elsarticle}

\usepackage{amssymb}
\usepackage{amsmath}
\usepackage{mathrsfs}

\journal{Journal of Magnetism and Magnetic Materials}

\begin{document}

\begin{frontmatter}



\title{The nontrivial ground state topology in the coexistence phase of
chiral d-wave superconductivity and 120 degrees magnetic order on a triangular lattice}


\author{V.\,V. Val'kov}
\ead{vvv@iph.krasn.ru}
\author{A.\,O. Zlotnikov}
\author{M.\,S. Shustin}
\address{Kirensky Institute of Physics, Federal Research Center KSC SB RAS, Krasnoyarsk, 660036 Russia}

\begin{abstract}
The $Z_ {2} $ topological invariant is defined in the chiral $d$-wave superconductor having a triangular lattice in the presence of the 120 degrees magnetic ordering. By analyzing the $Z_ {2}$ invariant, we determine the conditions of implementing nontrivial phases in the model with regard to superconducting pairings between nearest and next nearest neighbors. It is often supposed in such system that the pairing parameter between nearest neighbors should be equal to zero due to influence of the intersite Coulomb interaction. We show that taking into account even weak pairings in the first coordination sphere leads to the disappearance of the gapless excitations of the bulk spectrum in the wide region of the parameter space. Thus topological invariants can be defined in such region. Solving the problem of open edges it is shown that the zero energy modes are realized basically in the topologically nontrivial phases. Such zero modes are topologically protected Majorana modes. A connection between the $Z_ {2}$ invariant calculated at the symmetric points of the Brillouin zone with respect to the electron-hole symmetry and the integer topological invariant of the ground state of the 2D lattice expressed in terms of the Green functions is established in the presence of noncollinear magnetic ordering.
\end{abstract}


\end{frontmatter}


\section{Introduction}

Recently, much attention has been paid to the topological superconductors supporting Majorana zero modes. In pioneering works \cite{Read_Green-00, Kitaev-01}, such quasiparticles were predicted in the $p$-wave and effective $p$-wave superconductors. However, this type of superconductivity is still rather exotic for real materials. For the systems with $s$-wave pairing the several mechanisms have been proposed for the formation of the Majorana zero modes. One of the mechanisms
is characterized by the proximity-induced triplet $p_x + ip_y$ pairings on the surface layer of a
topological insulator in hybrid structures $s$-wave superconductor\,/\,topological insulator \cite{Fu_Kane-08}. Another mechanism is connected with  the combined influence
of strong spin-orbit interaction, proximity-induced superconductivity, and magnetic field
\cite{Sato-09, Sau-10, MacDonald-14}. In such case the Majorana zero modes arise when the external (or exchange)
magnetic field is greater than some critical field.

At present, the new mechanism of the formation of the Majorana edge states in topological spin-singlet
superconductors due to the presence of long-range magnetic order is often considered \cite{Martin-12, Wang-13, Gupta-15, Black-Schaffer-15}.
The symmetry of the superconducting state is considered to be chiral $d_{x^2-y^2}+id_{xy}$ supporting the non-trivial topology and edge states \cite{Volovik-97}. It should be noted that the time-reversal symmetry is broken in such state.
It is widely believed that the chiral $d$-wave superconductivity may be realized in materials with a
triangular lattice (for example, Na$_x$CoO$_2$ \cite{Wang-13}) and hexagonal lattice
(graphene \cite{Black-Schaffer-15}).

For the topological classification of the systems with many degrees of freedom as well as systems with strong electron
correlations the topological invariant $N_3$ expressed in terms of the Green functions has been derived \cite{Volovik-89, Volovik-03}. Using this invariant the quantum topological phase transitions have been studied in liquid helium $^3$He-B \cite{Volovik-09},
semiconducting nanowire \cite{Ghosh-10}, and quantum Hall systems.
It should be noticed, that in the systems with $2+1$-dimensions the $N_3$ topological invariant
is introduced for the gapped ground state \cite{Volovik-03}.

Non-zero values of $N_3$ indicate the non-trivial topology of the ground state supporting due to the bulk-boundary correspondence the edge states.
For 1D systems with the particle-hole symmetry the well-known $Z_{2}$
invariant (Majorana number) has been proposed \cite{Kitaev-01}.
Such invariant expressed in terms
of the Pfaffian of the Bogoliubov-de Gennes (BdG) Hamiltonian in the Majorana representation allows one to
study the conditions supporting the Majorana zero modes in systems with the gapped bulk excitation spectrum.
Later, the connection between $N_3$ and $Z_2$ numbers has been established
for the noncentrosymmetric superconductors with the broken time-reversal symmetry \cite{Ghosh-10}. The main result is that the Majorana zero modes is expected to appear in the states with odd $N_3$ invariant.

On a triangular lattice the appearance of the Majorana zero modes has been demonstrated in Ref. \cite{Wang-13} for the coexistence phase of $d_{x^2-y^2}+id_{xy}$-wave superconductivity and stripe non-collinear
magnetic ordering.
The superconducting pairings between nearest neighbors have been assumed to be suppressed by the inter-site Coulomb
interaction. Therefore the pairing interaction between the next nearest neighbors has been considered.

In \cite{Val'kov-16} on the basis of the self-consistent integral equations in the coexistence
phase for the $t-J-V$ model it has been shown that in the presence of stripe magnetic ordering
the superconducting order parameter does not have the chiral structure. Thus, the conditions for the realizations of the
Majorana zero modes on the triangular lattice have been analyzed for the coexistence phase of chiral superconductivity and 120$^{\circ}$
magnetic ordering \cite{Val'kov-16_2}. In Ref. \cite{Val'kov-16_2}, as well as in Ref. \cite{Wang-13},
the superconducting pairings in the second coordination sphere were only considered.
It turned out that the analysis of the topological phases in such model is
complicated due to the fact that there is a continual range of parameters for which
the bulk excitation spectrum is gapless. This is rather rare
since usually topological indices are introduced for a set of parameters in which the bulk excitation
spectrum is gapped. Therefore the edge states with the zero excitation energy have been found in the region with
the gapped bulk spectrum.

In this paper we study the conditions supporting the Majorana zero modes on the triangular lattice in the coexistence
phase of chiral $d$-wave superconductivity and 120$^{\circ}$ spin ordering with regard to the superconducting pairing in the second and first coordination spheres.
It is shown that taking into account the pairing between nearest neighbors with arbitrary small
amplitude $\Delta_{21}$ leads to the disappearance of the continuous parametric
region with the gapless bulk excitations. As the result the Majorana number and $N_3$ topological invariant
for a 2D lattice are calculated. The series of the topological phase transitions upon changing the chemical potential and exchange field is demonstrated.
The connection between $N_3$ and $Z_2$ invariants is determined in the presence of
noncollinear magnetism. Topologically non-trivial phases with the Majorana number
equal to -1 and odd $N_3$ invariant coincide with each other as well as with the
parameter regions supporting the Majorana zero modes which are found by solving the
problem with open boundary conditions.

\section{Model and method}

Let us consider the model describing the coexistence phase of chiral superconductivity
and noncollinear magnetic order in the mean-field approximation on the triangular lattice.
It is assumed that proximity-induced superconducting pairings appear between nearest and next-nearest neighbors leading to the $d_{x^2-y^2}+id_{xy}$-wave superconductivity.
This assumption allows us to study the topological phases of the system in the relatively simple
manner. However, it should be noted that the coexistence phase is caused not only by the proximity effect but also as a consequence of the internal electron interactions \cite{Val'kov-16}.

The long-range magnetic ordering is considered in the mean-field approximation assuming that an average magnetic moment $\left\langle {\bf S}_f \right\rangle = M \left( \cos({\bf QR}_f) -\sin({\bf QR}_f), 0 \right) $ is formed at the lattice site $f$. Here ${\bf{Q}}$ is the magnetic structure vector, $M$ is the average on-site magnetization. Hereinafter we consider the 120$^{\circ}$ spin ordering with ${\bf Q} = (Q, Q)$, $Q = 2\pi/3$ and define the coordinates in the real and quasimomentum space as ${\bf{R}}_{f}=n {\bf{a}}_{1} +m {\bf{a}}_{2}$, ${\bf{k}}=k_{1} {\bf{b}}_{1} +k_{2} {\bf{b}}_{2}$, where ${\bf{a}}_{i}$ and ${\bf{b}}_{i}$ are basic and reciprocal vectors of the triangular lattice, accordingly. The Hamiltonian has the form:
\begin{align}
\label{Ham_Tr_Lat_Real_Space}
H & = -\mu\sum_{f \sigma}c_{f \sigma}^{\dag}c_{f \sigma} +
\sum_{fm \sigma} t_{fm} c_{f \sigma}^{\dag}c_{m \sigma} +
\nonumber \\
& +  h \left(\bf{Q}\right) \sum_{f} \left( \exp({i\bf QR }_f) c_{f \uparrow}^{\dag}c_{f \downarrow} + \exp(-i{\bf Q R}_f) c_{f \downarrow}^{\dag}c_{f \uparrow}  \right) +
\nonumber \\
& + \sum_{fm} \left( \Delta_{fm}  c_{f \uparrow}c_{m \downarrow} + \Delta_{fm}^*  c_{m \downarrow}^{\dag}c_{f \uparrow}^{\dag}  \right),
\end{align}
where $\mu$ is the chemical potential, $t_{fm}$ and $\Delta_{fm}$ are the electron hopping and superconducting pairing amplitudes. The exchange field parameter is defined as follows:
\begin{equation}
\label{h_Thor}
h\left(\bf{Q}\right) = M/2 \sum_{m}I_{fm}\exp(-i {\bf Q}({\bf R}_f-{\bf R}_m)).
\end{equation}
$I_{fm}$ is the parameter of the exchange interaction, being considered within the two coordination spheres. An important difference of the system (\ref{Ham_Tr_Lat_Real_Space}) from the model studied in \cite{Val'kov-16_2} is the consideration of superconducting pairings between nearest neighbors.
Hereinafter, the model is considered both in the case of periodic boundary conditions along the direction ${\bf {a}}_{2}$ (the cylinder topology), and in the case of periodic boundary conditions in two spatial directions (the torus topology). In both cases, the operator part of the Hamiltonian (\ref{Ham_Tr_Lat_Real_Space}) has the form:
\begin{eqnarray}
\label{Ham_BdG}
H & = & \frac{1}{2}{\sum\limits_k {{{\bf C} }\left( k \right)}^+ } \cdot H\left( k \right) \cdot {\bf C }\left( k \right), \nonumber
\\
\label{Ham_kQ}
H(k) & = & \left( {\begin{array}{*{20}{c}}
{{{ \xi }_k}}&{h}&0&{{{ \Delta }_k}}\\
{{{ h}^ + }}&{{{ \xi }_{k - Q}}}&{ - \Delta _{ - k + Q}^T}&0\\
0&{ -  \Delta _{ - k + Q}^*}&{ -  \xi _{ - k + Q}^*}&{ - {{ h}^ + }}\\
{ \Delta _k^ + }&0&{ -  h}&{ -  \xi _{ - k}^*}
\end{array}} \right),
\end{eqnarray}
where ${\bf C}(k) =\left({\bf c}_{k\uparrow}, {\bf c}_{k-Q\downarrow}, {\bf c}^{+}_{-k+Q\uparrow},    {\bf c}^{+}_{-k\downarrow} \right)^{T}$.

In the case of the cylinder topology the operator ${\bf{C}}$ has  $4N_{1}$ components, where
$N_{1}$ is the number of sites along ${\bf{a}_{1}}$ direction.
Then, in the BdG Hamiltonian (\ref{Ham_BdG}) $N_{1}$ by $N_{1}$ matrices $\hat{\xi}_{k}$, $\hat{\Delta}_{k}$ and $\hat{h}$ have the form \textbf{($k \equiv k_2$)}:
\begin{eqnarray}
\label{Ham_Matrices}
\hat{\xi}_{k_{2}} &=& \left( {\begin{array}{*{20}{c}}
{{t_{k_{2}}} - \mu }&{T_{k_{2}}}&{\Gamma_{k_2}}&0&0\\
{T_{-k_{2}}^{}}& \ddots & \ddots & \ddots &0\\
{\Gamma_{-k_2}^{}}& \ddots & \ddots & \ddots &{\Gamma_{k_2}}\\
0& \ddots & \ddots & \ddots &{T_{k_{2}}}\\
0&0&{\Gamma_{-k_2}^{}}&{T_{-k_{2}}^{}}&{{t_{k_{2}}} - \mu }
\end{array}} \right),\nonumber
\end{eqnarray}
\begin{eqnarray}
\hat{h} &=& h \cdot diag\left( {e^{iQ},~e^{2iQ},\ldots ,~e^{N_{1}iQ}} \right),
\nonumber
\end{eqnarray}
\begin{eqnarray}
\hat{\Delta}_{k_{2}} &=&  - \left( {\begin{array}{*{20}{c}}
{\tilde \Delta_{k_{2}}^*}&{\psi_{ - {k_{2}}}^{*}}&{ \Delta _{22}^{*}{e^{ i{k_{2}}}}}&0&0\\
{\psi _{k_{2}}^*}& \ddots & \ddots & \ddots &0\\
{\Delta _{22}^{*}{e^{-i{k_{2}}}}}& \ddots & \ddots & \ddots &{\Delta _{22}^{*}{e^{ i{k_{2}}}}}\\
0& \ddots & \ddots & \ddots &{\psi _{ - {k_{2}}}^*}\\
0&0&{\Delta _{22}^{*}{e^{-i{k_{2}}}}}&{\psi _{k_{2}}^{*}}&{\tilde \Delta _{k_{2}}^{*}}
\end{array}} \right).\nonumber\\
\end{eqnarray}
Here
\begin{eqnarray}
\label{Func_of_matrices}
t_{k_{2}} &=& 2t_1\cos(k_{2}) + 2t_3\cos(2k_{2}),~~\tilde{\Delta}_{k_{2}} = 2\Delta_{21}\cos(k_{2}),~~~~\nonumber\\
{T_{k_{2}}} &=& {t_1}\left( 1 + \exp(ik_{2}) \right) + t_{2}\left(\exp(-ik_{2}) + \exp(2ik_{2}) \right),\nonumber\\
\Gamma_{k_2} &=& t_2\exp(ik_{2})+t_3\left(1+\exp(i2k_{2})\right), \nonumber\\
\Psi_{k_{2}} &=& \Delta_{22}\exp(i2\pi/3)\left(\exp(i2k_{2})+\exp(i2\pi/3-ik_{2}) \right) + \nonumber\\
&+& \Delta_{21}\exp(i2\pi/3)\left(1+\exp(i2\pi/3+ik_{2}) \right),
\nonumber
\end{eqnarray}
and $t_{1}$, $t_{2}$, $t_{3}$ are the hopping parameters for the first, second, and third coordination spheres. Parameters $\Delta_{21}$, $\Delta_{22}$ denote the amplitudes of the superconducting pairings of the $ d $-wave symmetry (angular momentum $ l = 2 $) which are implemented between the nearest and the next-nearest neighbors respectively.

The eigenvalues and eigenstates of the Hamiltonian (\ref{Ham_kQ}) determine the spectrum of elementary excitations as well as the amplitudes of the Bogoliubov quasiparticles:
\begin{eqnarray}
\label{alpha_kj}
\alpha_{k_2j} = \sum_{n=1}^{N_1} \left(A_{jn,k_{2}}c_{nk_2\uparrow}+B_{jn,k_{2}}c_{n,k_2-Q_2\downarrow}  +  \right.
\nonumber \\
\left. + C_{jn,k_{2}}c_{n,-k_2\downarrow}^{\dag}+D_{jn,k_{2}}c_{n,-k_2+Q_2\uparrow}^{\dag}\right).
\end{eqnarray}

When considering the lattice with the torus topology the value of $h$ is determined by the expression (\ref{h_Thor}), ${\xi}_{k} \equiv t_{{\bf k}}-\mu$, ${\Delta}_{k} \equiv \Delta_{{\bf k}}$, and functions $t_{\bf k}$, ${\Delta}_{\bf k}$ are the Fourier transforms of the hopping integral and the superconducting order parameter with  $d_{x^2-y^2}+id_{xy}$ symmetry type.
In should be noted that the pairing interaction in the first coordination sphere is considered to be sufficiently  suppressed by the inter-site Coulomb interaction, so $\Delta_{21}<<\Delta_{22}$ \cite{Wang-13, Val'kov-15}.

\section{Hamiltonian symmetry and $Z_{2}$ topological invariant}

Regardless of the consideration of the system with the cylinder or torus topology the BdG Hamiltonian (\ref{Ham_kQ}) has the symmetry:
\begin{eqnarray}
\label{Ham_Sym}
\Lambda H(k) \Lambda = -H^{*}(-k+Q),~~~~~
\Lambda = \left( {\begin{array}{*{20}{c}}
{\hat{0}}&\hat{I}\\
\hat{I}&{\hat{0}}
\end{array}} \right),
\end{eqnarray}
where $\hat{0}$ and $\hat{I}$ are the zeros and identity matrices of corresponding size ($2 \times 2$ for the torus topology and $2N_1 \times 2N_1$ for the cylinder topology). Due to this symmetry eigenvalues of the Hamiltonian $H(k)$
are grouped in pairs  $\varepsilon_{n}(k)$ and $-\varepsilon_{n}(-k+Q)$.
Following the paper \cite{Ghosh-10}, let us consider the particle-hole invariant momenta $K$ (PHIM points) of the Brillouin zone such that $K = -K+Q+G$, where $G$ is a reciprocal-lattice vector. At these points the BdG Hamiltonian has the particle-hole symmetry. In the case of the cylinder topology $K_{2} = -2\pi/3;~\pi/3$, while in the case of the torus topology we have four PHIM points ${\bf{K}} = (-2\pi/3,~-2\pi/3);~(-2\pi/3,~\pi/3);~(\pi/3,~-2\pi/3);~(\pi/3,~\pi/3)$.
Then we can define matrices
\begin{eqnarray*}
\label{W_Matrix}
W(k) & = & H(k)\Lambda, \, \, \, \tilde{W}(k) = R^{T}W(k)R,
\nonumber \\
R & = & \frac{1}{\sqrt{2}}\left( {\begin{array}{*{20}{c}}
{\hat{I}}&-i\hat{I}\\
\hat{I}&{i\hat{I}}
\end{array}} \right),
\end{eqnarray*}
which satisfy the relations $W(k) = -W^{T}(-k+Q)$, $\tilde{W}(k) = -\tilde{W}^{T}(-k+Q)$.
These matrices are the antisymmetric ones at the PHIM points. It can be shown that the matrix $\tilde{W}$ coincides with the BdG Hamiltonian (\ref{Ham_kQ}) if we will use in the expression (\ref{Ham_BdG}) the Majorana representation:
\begin{eqnarray}
c_{k\sigma} = \gamma_{Ak\sigma} + i\gamma_{Bk\sigma};~~c^{+}_{k\sigma} = \gamma_{A-k\sigma} - i\gamma_{B-k\sigma}.
\end{eqnarray}

Thus, following Kitaev \cite{Kitaev-01}, one can introduce the $Z_{2}$ Pfaffian invariant $M(K_{2})$ in the cylinder topology only for the PHIM points $K_{2}=-2\pi/3,~\pi/3$:
\begin{eqnarray}
\label{Z2_M}
M(K_{2})=P(K_{2}, K_{1}=-2\pi/3)P(K_{2},K_{1}=\pi/3),
\end{eqnarray}
where $P({\bf K})$ is the fermionic parity of the ground state of the system with the torus topology:
\begin{eqnarray}
\label{Par}
P({ \bf K})=sign\left(Pf\left( \tilde{W}(\bf{K}) \right)\right).
\end{eqnarray}
If $M(K_{2})=-1$ the system is in the topologically nontrivial phase supporting the Majorana zero modes. Otherwise, if $M(K_{2})=1$ the ground state is topologically trivial and there is no topologically protected edge states with zero excitation energy. Note that the bulk spectrum should be gapped to define the Majorana number.
In general $P({{\bf{K}}})=sign\left( {h^2 - \xi_{\bf{K}}^{2} - |\Delta_{\bf K}|^2}\right)$ and it can be shown by the direct calculations that $P(K_{2}=\pi/3,K_{1}=-2\pi/3)=P(K_{2}=\pi/3,K_{1}=\pi/3)$. It means that at $K_{2}=\pi/3$  there is no the Majorana zero modes in the system considering the cylinder topology.

\begin{figure}[htb!]
\begin{center}
\includegraphics[width=0.4\textwidth]{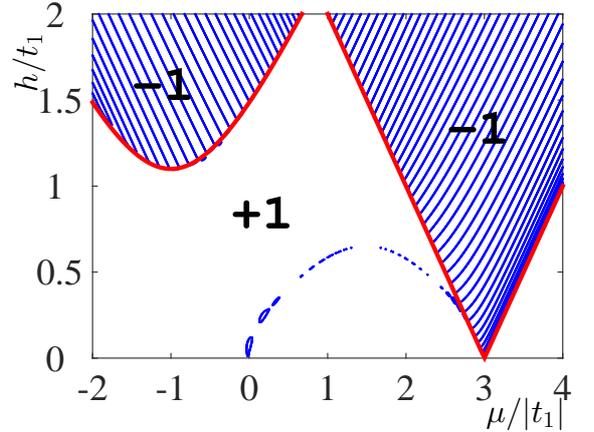}
\caption{The conditions for the realization of the zero energy excitations obtained with consideration of the periodic boundary conditions along ${\bf {a}}_{2}$ (blue thin lines), $h$ is the exchange field, $\mu$ is the chemical potential, $t_1$ is the hopping parameter between nearest neighbors. The parameters are $t_{1}=-1$, $t_{2}=t_{3}=0$, $\Delta_{22} = 0.3 |t_1|$, $\Delta_{21} = 0.05|t_{1}|$, $N_1 = 48$.
The red bold lines show a border between the phases with different values of the topological $Z_{2}$ invariant $M = \pm 1$.}
\end{center}
\label{Lines_N48}
\end{figure}

At $K_{2} = -2\pi/3$ the Majorana number is defined by the relation
\begin{eqnarray}
\label{M_non_trivial}
M&=&sign\Big(\left(h^2 - (\mu + 3t_{1} - 6t_{2} + 3t_{3})^2\right)\cdot \nonumber\\
&\cdot&\left( h^2 - (t_{1}-t_{2}-3t_{3}-\mu)^2 - 4\left(\Delta_{21} - 2\Delta_{22} \right)^2\right)\Big).
\nonumber \\
\end{eqnarray}

In Figure 1 the parameters for which the Majorana number (\ref{M_non_trivial}) change sign are depicted by the bold lines. As it will be shown in the last paragraph for these parameters the bulk excitation spectrum becomes gapless.  The curves $h(\mu)$ for which the gapless elementary excitations occur on the triangular lattice with the cylinder topology and $N_{1} = 48$ are shown by thin lines. It can be seen that the majority of the zero modes lies in the topologically nontrivial phase. As $N_{1}$ increases the distribution of the zero energy curves becomes dense and all of them are found in the topologically nontrivial phases with $M = -1$. The more $N_{1}$ decreases, the more zero modes exist in the topologically trivial phase. These modes are not topologically protected. This indicates that the correspondence between bulk and boundary is well established when considering a sufficiently large number of sites. The set of parameters is chosen as $\Delta_{21}=0.05|t_{1}|$, $\Delta_{11} = 0.3|t_{1}|$, $t_{2}=t_{3}=0$.

\begin{figure}[htb!]
\begin{center}
\includegraphics[width=0.45\textwidth]{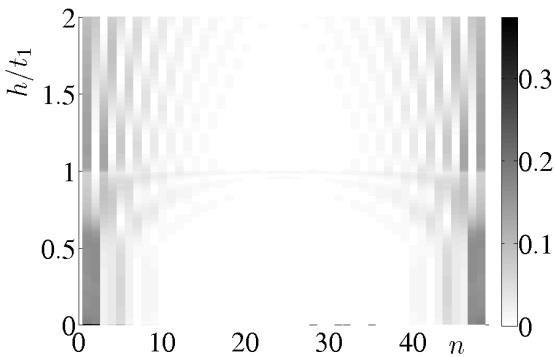}
\caption{Spatial distribution of the sum of the Bogoliubov coefficients $p_{n}(h)$ (color bar) vs exchange field $h$ at $\mu/|t_{1}|=2$. The other parameters are the same as in Figure~1. The darkest and lightest
areas correspond to the largest and smallest values of $p_{n}$, respectively.}
\end{center}
\label{UV_mu_eq2}
\end{figure}
\begin{figure}[htb!]
\begin{center}
\includegraphics[width=0.45\textwidth]{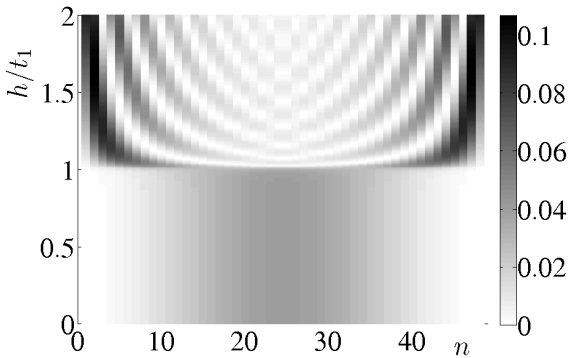}
\caption{Spatial distribution of the sum of the Bogoliubov coefficients
$p_{n}(h)$ (color bar) vs
exchange field $h$ at $\mu/|t_{1}|=4$. The other parameters are the same as in Figure~1. The darkest and lightest
areas correspond to the largest and smallest values of $p_{n}$, respectively.}
\end{center}
\label{UV_mu4}
\end{figure}

It should be noted that changing the parameters in the topologically nontrivial phase leads to the oscillations of the minimal excitation energy $\varepsilon_{0}$ and its dropping to zero on the lines of the zero modes (thin lines in Figure~1). At the points where $\varepsilon_{0} = 0$ a transition is realized: the ground state containing a superposition of states with an even number of fermions is replaced by a state with an odd number of fermions and vice versa.
Such switching of the fermionic parity has been obtained for the Kitaev model \cite{Hedge-16} and probably it is a general property of the finite quasi-one-dimensional systems in the topologically nontrivial phase.

It is sufficient that all of the zero modes shown in Figure~1 are the edge ones. This is an important difference from the case $\Delta_ {21} = 0$ considered in \cite{Val'kov-16_2} where the continual region with the bulk gapless excitations is appeared in the space of the parameters $h$ and $\mu$. As the result the zero energy modes of the system with the cylinder topology which are found in this region are not the edge ones and represent the bulk excitations modified due to the boundary effects. With regard to the weak nearest neighbors superconducting pairing $\Delta_{21} << \Delta_{22}$ all of the zero modes in the topologically nontrivial phases become the edge ones.

Let us demonstrate the realization of the edge states for $\Delta_{21} \neq 0$ in the topological phase studying the dependence of the site-dependent parameter
\begin{eqnarray}
\label{p_vs_h}
p_{n}(h)=|A_{0n,K_{2}}|^2 + |B_{0n,K_{2}}|^2 + |C_{0n,K_{2}}|^2 + |D_{0n,K_{2}}|^2
\end{eqnarray}
on the exchange field $h$ (vertical panel) and the number of site $n$ (horizontal panel).
The Bogoliubov coefficients appearing in (\ref{p_vs_h}) correspond to an elementary excitation with a minimal energy $\varepsilon_ {0}$ and $K_{2}=-2\pi/3$. In Figures 2,~3 such dependencies are shown for $\mu=2|t_{1}|$ and $\mu=4|t_{1}|$. The other parameters are the same as in Figure~1. In these cases the transition between the topologically trivial and nontrivial phases corresponds to $h=|t_1|$.  It is seen in Figure~3 that the edge states including the Majorana zero modes are realized in the phase with $M=-1$ of the $Z_{2}$ topological invariant (\ref{Z2_M}). In Figure~2 the edge states and zero modes are found even in the phase with $M=1$. As it will be shown in the next paragraph the edge states can exist in this region but the zero modes are not topologically protected.
Upon increasing $N_{1}$ the Majorana zero modes become more localized at the edges.

The following two features deserve mention. First, the topologically protected edge states with non-zero excitation energy can be realized even if the value of the $Z_{2}$ invariant corresponds to  the topologically trivial phase ($M=1$, see Figure~2). This result is in agreement with the calculation of the $N_3$ invariant for a 2D system considered below. Second, in the considered system the edge states with zero energy can be realized with quasimomenta $k_{2} \ne -2\pi/3$ but such states are not topologically protected ones.

\section{The topological invariant $N_3$ of a 2D lattice and its connection with the $Z_{2}$ invariant. The analysis of the bulk spectrum }

It is known that topological transitions changing the topological index occur when the gap closes in
the bulk spectrum. For the system under consideration, the bulk spectrum has the form:
\begin{eqnarray}
\label{bulk_spectrum}
E_{\bf{k}}^ \pm  = \sqrt {\frac{1}{2}\left( \xi _{\bf{k}}^2 + \xi _{\bf{k} - \bf{Q}}^2 + 2{h^2} + \left| \Delta _{\bf{k}} \right|^2 + \left| \Delta_{\bf{-k+Q}} \right|^2  \right) \pm v_{\bf{k}}^2},
\end{eqnarray}
where
\begin{eqnarray}
\nu_{\bf{k}}^2 = \left\{ \frac{1}{4} \left( \xi_{\bf{k}}^2 - \xi_{\bf{k-Q}}^2 + |\Delta_{\bf{k}}|^2 - |\Delta_{\bf{-k+Q}}|^2  \right) + \right.
\nonumber \\
\left. + h^2 \left[ \left( \xi_{\bf{k}} + \xi_{\bf{k-Q}} \right)^2 + \left|\Delta_{\bf{k}}+\Delta_{\bf{-k+Q}}\right|^2 \right] \right\}^{1/2}.
\end{eqnarray}
The conditions for zero energy in the bulk spectrum are described by the equation:
\begin{eqnarray*}
\label{bulk_gap_close}
|h^2 - \xi_{\bf{k}}\xi_{\bf{k-Q}} - \Delta_{\bf{k}}\Delta_{\bf{-k+Q}}^{*}|^{2} + |\xi_{\bf{k}}\Delta_{\bf{-k+Q}} - \xi_{\bf{k-Q}}\Delta_{\bf{k}}|^2 = 0.
\end{eqnarray*}
At the PHIM points $\bf{K}=\bf{-K+Q+G}$ the second term in the equation is identically equal to zero, and the first term is the same as $Pf(\tilde{W(\bf{K})})$. Thus, at the symmetric points of the Brillouin zone the change in the sign of the Majorana number (\ref{Z2_M}), as it should be, is accompanied by the existence of zero energy in the bulk spectrum at these points.

In the case when $\bf{k} \neq \bf{-k + Q}$, the equations determining the conditions for the realization of the gapless bulk excitations have the form:
\begin{eqnarray}
\label{bulk_gap_eqns}
h^2 - \xi_{\bf{k}}\xi_{\bf{k-Q}} - \Delta_{\bf{k}}\Delta_{\bf{-k+Q}}^{*}&=&0, \nonumber\\
|\xi_{\bf{k}}\Delta_{\bf{-k+Q}} - \xi_{\bf{k-Q}}\Delta_{\bf{k}}|&=&0, \nonumber\\
\text{Im}(\Delta_{\bf{k}}\Delta_{\bf{-k+Q}}^{*}) &=&0.
\end{eqnarray}

The formation of the gapless bulk excitations at the non-PHIM points according to the solution of the Eqs. (\ref{bulk_gap_eqns}) also leads to a topological phase transitions. However, at this transition the $Z_ {2}$ invariant (\ref {Z2_M}) does not change. A characteristic that allows to identify such transitions in two-dimensional systems (including the systems with interaction) is the topological invariant of the ground state introduced in Ref. \cite{Volovik-89}:
\begin{eqnarray}
\label{Z_inv}
N_{3} &=& \frac{1}{24\pi^{2}}\varepsilon_{\mu\nu\lambda}\int_{-\infty}^{\infty}d\omega\int_{-\pi}^{\pi}
\int_{-\pi}^{\pi}dk_{1}dk_{2}Sp\Big(G\partial_{\mu}G^{-1}\cdot \nonumber\\
&\cdot& G\partial_{\nu}G^{-1}G\partial_{\lambda}G^{-1} \Big),
\end{eqnarray}
where $\mu,~\nu,~\lambda = 1,2,3$, $\varepsilon_{\mu\nu\lambda}$ is antisymmetric Levi-Civita tensor,
$\partial_{1}=\partial/\partial_{k_1}$, $\partial_{2}=\partial/\partial_{k_2}$, $\partial_{3}=\partial/\partial_{\omega}$. By repeated indices we mean the summation. In the system of noninteracting electrons the matrix Green function $G$ is a matrix $4 \times 4$ for the torus topology and has the form $G = \big[ i\omega I - H(\bf{k}) \big]^ {-1}$.

Non-zero integer values of the invariant $N_3$ (\ref{Z_inv}) determine the topologically nontrivial phases in which edge states can form. In the Ref. \cite{Ghosh-10} a connection between the $Z_ {2}$ invariant (\ref{Z2_M}) and the topological invariant $N_3$ (\ref{Z_inv}) was established in the case of noncentrosymmetric systems with the broken time-reversal symmetry but preserving the electron-hole symmetry. It was shown that the product of the Majorana numbers (\ref{Z2_M}) at the points $ { \bf k} = -{\bf k} + {\bf G} $ coincides with the parity of the topological index $ N_ {3}$. In a system with magnetic ordering this relation is generalized:
\begin{eqnarray}
\label{Z_Z2_connect}
(-1)^{N_{3}} = sign\Big( \prod_{\bf{K}=-\bf{K+Q+G}} Pf\left(\tilde{W}(\bf{K})\right) \Big).
\end{eqnarray}

\begin{figure}[htb!]
\begin{center}
\includegraphics[width=0.45\textwidth]{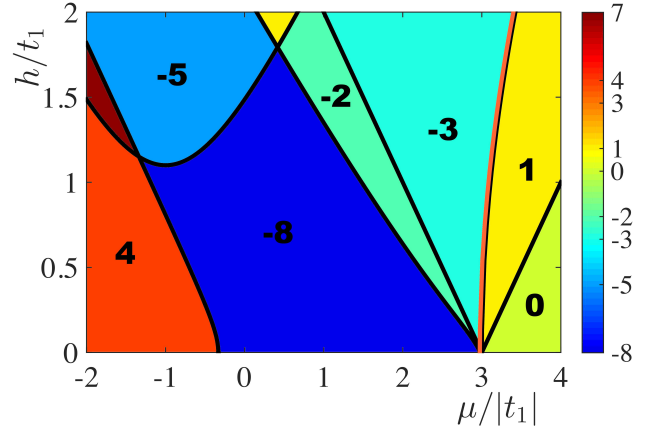}
\caption{The diagram of topological phases with different $N_{3}$ (\ref{Z_inv}) in the variables $h$, $\mu$, where $h$ is the exchange field, $\mu$ is the chemical potential. The parameters are the same as in Figure 1. The phases with an odd $N_ {3}$ value correspond to the phases with the Majorana zero modes in Figure 1 having $M = -1$ (\ref{Z2_M}).}
\end{center}
\label{N3_map}
\end{figure}

The phase diagram with different topological phases in the space of the chemical potential $\mu$ and the exchange field $h$ is shown in Figure 4. In each phase the values of the topological invariant $ N_3 $ are marked. The solid lines defining the boundaries between different topological phases are obtained as the solutions of the system of equations \eqref{bulk_gap_eqns} which are determined the presence of the bulk gapless excitations. The parameters are the same as in Figure 1. It should be noted that this invariant is ill-defined at the topological transition point. In the vicinity of the transition the calculation of the invariant requires an increase in the accuracy. As can be seen from Figure 4 the increase of the chemical potential leads to a series of topological transitions. The topological trivial phase  with $N_3 = 0$ at  $ \mu < -2|t_1|$ is implemented when the chemical potential intersects the bottom of the bare electron band (not shown in the Figure).

As is well known, the difference between the values of the $N_ {3}$ invariant in neighboring phases determines the values of the topological invariants of the Fermi points in which the bulk spectrum has zero energy at the transition between phases. In the model under consideration the invariants of the Fermi points are equal to $\pm 1$. There is only one exception at the transition between phase with $N_3 = 0$ to the phase with $N_3 = 4$ at negative $\mu$ (not shown in Figure 4) where the invariant of each from two Fermi points is 2.  Thus at other cases, the difference corresponds to a number of the nodal points of the bulk spectrum at the topological transition.

The excitation spectrum in the coexistence phase of superconductivity and noncollinear magnetic order differ from the spectrum in a superconducting phase.
Moreover, the spectrum in the coexistence phase is determined by the two superconducting order parameters $\Delta_{{\bf k}}$, $ \Delta_{{\bf -k+Q}}$, which have the different systems of the nodal points. This leads to several significant differences in the analysis of the zeros of the bulk spectrum. The spectrum in superconductors has zero energy at the boundaries and in the middle of the Brillouin zone only when the chemical potential intersects the bottom or top of the bare electron band. In the coexistence phase due to the exchange field the spectrum has zero energy at these points, when the chemical potential lies inside the band. Such picture can be seen from Figure 4 at the transition from the phase with $ N_3 = -2 $ to the phase with $ N_3 = -3 $ when the gap closes at the point $(-2\pi/3,-2\pi/3)$ under the condition $h = |\mu + 3t_1-6t_2 + 3t_3|$. This point corresponds to the one of the nonequivalent points lying at the intersection of the edges of the hexagonal Brillouin zone. In this case $\Delta_{{\bf k}} = \Delta_{{\bf -k+Q}} = 0$. The second analogous transition is realized between the phases with $N_3 = 3$ and $N_3 = 1$ when the spectrum becomes gapless at the points $(0,0 $ and $(2\pi/3,2\pi/3)$. At small values of $\Delta_{21}$ the phase with $N_3 = 3$ is rather narrow and lies between the phases with $N_3 = -3$ and $N_3 = 1$. This narrow phase is schematically shown in Figure 4. For a superconductor without magnetic ordering the intersection of the nodal points of the superconducting order parameter by the Fermi contour leads to the gapless excitations. When noncollinear magnetism is taken into account this condition is not satisfied due to the parameter $\Delta_{{\bf -k+Q}}$. However, there are conditions when the energy spectrum is equal to zero at the points in which  $\Delta_{{\bf k}}, \Delta_{{\bf -k+Q}} \ne 0$. This picture corresponds to the remaining transitions in Figure 4. It should be noted that with disregard to $\Delta_{21} $ the relation $\Delta_{{\bf k}} = \Delta_{{\bf -k+Q}}$ is satisfied, as a result the energy spectrum is considerably simplified. When $\Delta_ {21}$ is taken into account the condition $\Delta_{{\bf k}} = \Delta_{{\bf -k+Q}} \ne 0 $ is valid only at the points $(-2\pi/3,\pi/3)$, $(\pi/3,-2\pi/3)$, $(\pi/3,\pi/3)$. The zeros of the spectrum at these points are realized, for example, at the transition from the phase with $ N_3 = -8$ to the phase with $ N_3 = -5$ upon increasing $h$.

From Eq. (\ref{Z_Z2_connect}) we conclude that the Majorana modes exist in the phases with odd $N_ {3}$. The transition to such phases is accompanied by the closing the gap in the bulk spectrum in an odd number of points in the Brillouin zone. In the phases with even $N_ {3}$ the edge states can arise but the topologically protected zero modes are not found. This agrees with the calculation results shown in Figures 1-3.
These conclusions indicate that the definition of the topological invariant (\ref{Z_inv}) allows one to search for possible conditions for the realization of the Majorana modes in electron systems with interaction and magnetic order.

\section{Conclusions}

The topological properties of the coexistence phase of the $d_{x^2-y^2}+id_{xy}$-wave superconductivity and the noncollinear 120$^{\circ}$ magnetic ordering on a triangular lattice are studied. When the superconducting pairings are taken into account only in the second coordination sphere, the gap in the bulk excitation spectrum is closed in the continuous region of the parameter space. This feature means that the topological invariants cannot be introduced in a standard way, in spite of the fact that the edge zero modes are found in the system. In the present work it is shown that taking into account the arbitrarily small superconducting amplitude, induced by the pairing interaction in the first coordination sphere, leads to opening a gap in the bulk spectrum. The bulk spectrum becomes gapless only on the boundaries between topologically different phases. This allows us to introduce the $Z_{2}$ topological invariant $M$ (the Majorana number) and analytically determine the conditions of realizing the topologically nontrivial phases with $M = -1$.

Considering the triangular lattice with periodic boundary conditions along the basic vector ${\bf{a}}_ {2}$, the zero modes are found to exist on the specific curves in the parameter space of the chemical potential and exchange field. It is shown that for the lattice with a finite number of sites $N_ {1}$ along the direction ${\bf{a}}_ {1}$ the zero modes can arise in the topologically trivial phase. Such zero modes are not topologically protected. However the majority of the zero modes (the Majorana modes) lies with increasing $N_ {1}$ in the topologically nontrivial phase with $ M = -1 $.

The topological invariant $N_3$ of the 2D lattice expressed in terms of the Green functions is calculated for the coexistence phase. We find a series of the topological transitions in the coexistence phase upon increasing the chemical potential. A relationship between the two topological invariant $M$ and $N_{3}$ is determined with regard to the noncollinear magnetism. It is shown that topologically nontrivial phases with the Majorana number equal to $-1$ correspond to the phases with odd $N_3$. In the topologically nontrivial phases with even $N_3$ the edge states can exist but they cannot be the topologically protected Majorana edge states with zero excitation energy.

\section{Acknowledgments}

This study was funded by the Russian Foundation for Basic Research, Government of Krasnoyarsk Territory, and Krasnoyarsk Region
Science and Techonology Support Fund according to the research projects Nos. 16-02-00073-a (calculation of the topological invariant of a 2D lattice) and 16-42-243069-mol-a (calculation of the $Z_2$ invariant). A.O.Z. is grateful for support of the Grant of the President of the Russian Federation SP-1370.2015.5. The work of M.S.S.
was supported by the Grant of the President of the Russian Federation MK-1398.2017.2.



%% \newpage




%%\bibitem{}

%%\end{thebibliography}

\end{document}